\documentclass[headings=standardclasses]{scrartcl}

\usepackage{geometry}
\usepackage[utf8]{inputenc}
\usepackage[T1]{fontenc}
\usepackage[onehalfspacing]{setspace}
\usepackage{charter}
\usepackage[charter]{mathdesign}
\usepackage{microtype}
\usepackage{amsmath}
\usepackage{amsthm}
\usepackage{tabulary}
\usepackage{longtable}
\usepackage{booktabs} 
\usepackage{diagbox}
\usepackage{ragged2e}

\usepackage{siunitx}
\usepackage{bm}
\usepackage[pdftex]{graphicx}
\usepackage{rotating}
\usepackage{pdfpages}
\usepackage{pdflscape}
\usepackage{color}
\usepackage{array}
\usepackage[font=bf]{caption}
\usepackage{subcaption}
\usepackage{pgfplots}
\usepackage{pgfplotstable}
\usepackage{float}
\usepackage[title]{appendix}
\pgfplotsset{compat=1.9}
\usepackage{url}
\usepackage[section]{placeins}
\usepackage{natbib}
\usepackage{hyperref}
\hypersetup{
	colorlinks   = true,    
	urlcolor     = blue,    
	linkcolor    = blue,    
	citecolor    = blue      
}

\usepackage[boxed]{algorithm2e}
\SetKwFunction{KwInit}{Initialisation}
\SetKwComment{Comment}{\% }{}

\makeatletter
\renewcommand*\env@matrix[1][c]{\hskip -\arraycolsep
  \let\@ifnextchar\new@ifnextchar
  \array{*\c@MaxMatrixCols #1}}
\makeatother

\usepackage{xcolor}
\usepackage{soul}

\usepackage{verbatim}

\usepackage[english]{babel}
\usepackage{changepage}
\usepackage{graphicx}
\usepackage{enumerate}
\usepackage{tabularx,longtable,supertabular,booktabs,dcolumn,multicol,multirow}
\usepackage{rotating}

\begin{document}
\thispagestyle{empty}
\begin{spacing}{1.2}
\begin{flushleft}
\huge \textbf{Designed Quadrature to Approximate Integrals in Maximum Simulated Likelihood Estimation} \\
\vspace{\baselineskip}
\normalsize
October 31, 2020 \\
\vspace{\baselineskip}
\textsc{Prateek Bansal} (corresponding author)\\
Transport Strategy Centre, Department of Civil and Environmental Engineering\\
Imperial College London, UK \\
prateek.bansal@imperial.ac.uk \\
\vspace{\baselineskip}
\textsc{Vahid Keshavarzzadeh} \\
Scientific Computing and Imaging Institute \\ 
University of Utah \\
vkeshava@sci.utah.edu\\
\vspace{\baselineskip}
\textsc{Angelo Guevara} \\
Departamento de Ingenier\'{i}a Civil\\
Universidad de Chile \\
crguevar@ing.uchile.cl \\
\vspace{\baselineskip}
\textsc{Ricardo A. Daziano} \\
School of Civil and Environmental Engineering \\
Cornell University \\
daziano@cornell.edu \\
\vspace{\baselineskip} 
\textsc{Shanjun Li}\\
Dyson School of Applied Economics and Management \\
Cornell University \\
sl2448@cornell.edu \\
\end{flushleft}
\end{spacing}

\newpage
\thispagestyle{empty}
\section*{Abstract}
Maximum simulated likelihood estimation of mixed multinomial logit (MMNL) or probit models requires evaluation of a multidimensional integral. Quasi-Monte Carlo (QMC) methods such as shuffled and scrambled Halton sequences and modified Latin hypercube sampling (MLHS) are workhorse methods for integral approximation. A few earlier studies explored the potential of sparse grid quadrature (SGQ), but this approximation suffers from negative weights. As an alternative to QMC and SGQ, we looked into the recently developed designed quadrature (DQ) method. DQ requires fewer nodes to get the same level of accuracy as of QMC and SGQ, is as easy to implement, ensures positivity of weights, and can be created on any general polynomial spaces. We benchmarked DQ against QMC in a Monte Carlo study under different data generating processes with a varying number of random parameters (3, 5, and 10) and variance-covariance structures (diagonal and full). Whereas DQ significantly outperformed QMC in the diagonal variance-covariance scenario, it could also achieve a better model fit and recover true parameters with fewer nodes (i.e., relatively lower computation time) in the full variance-covariance scenario. Finally, we evaluated the performance of DQ in a case study to understand preferences for mobility-on-demand services in New York City. In estimating MMNL with five random parameters, DQ achieved better fit and statistical significance of parameters with just 200 nodes as compared to 1000 QMC draws, making DQ around five times faster than QMC methods.\\

\textit{Keywords:} Monte Carlo Integration; Designed Quadrature; Sparse Grid Quadrature; Quasi Monte Carlo; Mixed Logit 

\newpage

\section{Introduction}
Discrete choice models are widely applied across several disciplines such as marketing, economics, and travel behavior. The mixed multinomial logit (MMNL) model currently dominates empirical choice modeling research since it can capture unobserved preference heterogeneity in willingness to pay (WTP) of decision-makers. The multinomial probit (MNP) model is also an attractive alternative to specify flexible substitution patterns across alternatives, as well as to jointly model mixed types of dependent variables \citep{bhat2015new}. In the maximum likelihood estimator of both MMNL and MNP models, choice probabilities involve computation of a multidimensional integral \citep{train2009discrete}. Moreover, estimating design criteria in Bayesian D-efficient designs of choice experiments also requires computation of multidimensional integrals \citep{yu2010comparing}.

Among a handful of analytic solutions, a simulation-free maximum approximate composite marginal likelihood (MACML) estimation approach is available for MNP models \citep{bhat2011maximum}, but the Geweke - Hajivassiliou - Keane (GHK) simulator \citep{geweke1994alternative} still is more commonly used in practice. In the absence of a tractable analytical solution, these integrals are generally approximated through simulation in the estimation of logit models and creating Bayesian D-efficient designs. In general, the above-mentioned estimation problems include evaluation of integrals of the following type: 
\begin{equation*}
  \int_\Gamma f(\bm{x}) \omega(\bm{x}) d\bm{x} \approx \sum_{q=1}^n f(\bm{x}_q) w_q,
\end{equation*}
where $\Gamma$ is a set in the $d$-dimensional Euclidean space $\mathbb{R}^d$, $\omega$ is a probability density function (or positive weight function), and $f(.)$ is generally a conditional likelihood function. Instead of solving the actual integral, simulation-based inference considers a discrete approximation. The objective of computationally efficient simulation is to determine nodes $\bm{x}_q$ and weights $w_q$ so that integration can be approximated with minimum number of function evaluations ($n$).   

Simulation-based inference in discrete choice models started with Pseudo-Monte Carlo (PMC) methods. As an alternative to PMC, Quasi-Monte Carlo (QMC) methods are now typically used to approximate multidimensional integrals \citep{bhat2001quasi,train2009discrete}. More specifically, low-discrepancy sequences\footnote{\cite{dick2014discrepancy} illustrates that the lower the discrepancy of a sequence, the smaller will be the error in the Monte Carlo integration.} such as randomized and scrambled Halton sequences \citep{bhat2003simulation} and modified latin hypercube sampling (MLHS) \citep{hess2006use} dominate the empirical literature. QMC methods are preferred over PMC because QMC requires fewer draws (i.e., fewer Loglikelihood function counts) to approximate the integrals due to their excellent coverage properties \citep{bhat2001quasi}. \cite{sandor2004quasi} and \cite{munger2012estimation} showed superiority of \emph{digital nets} over Halton sequences, but implementation simplicity of the latter makes it a popular alternative in practice. 

Empirical instability of point estimates with a low number of evaluations when using either PMC or QMC has motivated researchers to explore easy-to-implement numerical methods that can accurately approximate the integral of interest with fewer function (i.e., integrand) evaluations ($n$) than QMC. In this study, we argue and illustrate that recent developments in quadrature methods open such possibilities. 

\subsection{Quadrature Methods and Research Gap}
As an alternative to QMC, quadrature methods have been explored in the discrete choice literature  \citep{heiss2008likelihood,heiss2010panel,abay2015evaluating, patil2017simulation, goos2018quadrature}. Quadrature methods mainly differ from QMC in two ways, as quadrature a) generally assumes that the integrand can be approximated on a polynomial space; b) uses deterministic draws (or \textit{nodes}) that carry unequal weights.

The Gaussian quadrature method approximates one-dimensional integrals with just a few nodes.\footnote{A $K-$times differentiable integrand can be approximated by a polynomial of degree $K$, and thus the resulting integral with surrogate integrand can be approximated using just $\frac{K+1}{2}$ nodes \citep{golub1969calculation}.} Quadrature can be simply extended to multiple dimensions using the tensor product. However, this multidimensional extension of quadrature suffers from the curse of dimensionality -- the number of nodes (i.e., function evaluations) increases exponentially with the number of dimensions, making it impractical beyond 4-5 dimensions. \cite{smolyak1963quadrature} proposed a way to extend the univariate quadrature rule to multiple dimensions in a method that is often called sparse grid quadrature (SGQ) in the literature. For example, whereas Gaussian quadrature can exactly compute an integral with a univariate polynomial of order 5 with 3 nodes, the same function in 20 dimensions requires $3^{20} =3,486,784,401$ nodes in product rule quadrature and $841$ nodes in SGQ, respectively \citep{heiss2008likelihood}.   

\cite{heiss2008likelihood} have demonstrated that SGQ performs much better than QMC in estimation of the MMNL model with even up to 20 random parameters. Further, \cite{heiss2010panel} combined SGQ with efficient importance sampler (EIS)\citep{richard2007efficient} to estimate MNP and panel binary probit models, and demonstrated superiority of this hybrid SGQ-EIS approach over traditional QMC methods. 

Even if nodes and weights in SGQ can be pre-computed and stored for reuse as easily as in traditional QMC methods, SQG methods have not been adopted in practice due to three possible reasons. \textit{First}, weights computed in SGQ can be negative. Whereas \cite{heiss2008likelihood} discussed this concern as an eventual possibility, they claimed not to encounter any such issue -- perhaps due to a very simplistic simulation design with a (low-variance) diagonal variance-covariance matrix. In contrast, in our experience we always encountered the issue of negative choice probability estimates for a few individuals coming from negative weights, which numerically led to imaginary (complex) loglikelihood values. \cite{patil2017simulation} also encountered convergence issues due to negative weights while applying the SGQ-EIS method in estimation of multinomial probit. \textit{Second}, the required number of nodes to accurately approximate the integral using SGQ depends on the functional properties of the integrand, but the researcher is generally not aware of these properties. \textit{Third}, whereas SGQ reduces the number of nodes significantly as compared to the product rule, the cardinality remains very high relative to that of QMC for high dimensional integrals. Concerns two and three can be illustrated with the following example. If the integrand in a ten-dimensional integral can be well approximated using a $3^{rd}$-order polynomial, Gaussian SGQ just needs 21 nodes, but the number of required nodes and thus the number of function evaluations increases to $8,761$ for a $9^{th}$-order polynomial \citep{heiss2008likelihood}. The combined consequences of concerns two and three is confirmed by \cite{abay2015evaluating} in estimation of a panel binary probit -- SGQ outperforms QMC for dimensions below or equal to 4, but QMC starts dominating SGQ for higher dimensions, and the difference is apparent as panel covariance increases. This is because a higher panel covariance in binary probit makes the integrand (i.e., loglikelihood) less smooth and therefore, a higher order polynomial (i.e., higher number of nodes or function evaluations) are required to approximate the integral at the same level of accuracy.  

\subsection{Moment-base Quadrature and Contributions}
More recent developments in quadrature methods could address the main concerns of SGQ. Whereas \cite{ryu2015extensions} showed that numerical quadrature can be obtained by solving an infinite-dimensional linear program (LP), \cite{jakeman2018generation} used the same flexible moment-based optimization framework to obtain a numerical quadrature rule. Very recently, \cite{keshavarzzadeh2018numerical} simplified this moment-based strategy by solving a relaxed version of the original optimization problem and came up with a new numerical quadrature rule known as designed quadrature (DQ). 

DQ has many key features. This flexible framework allows the researcher to add a constraint to ensure  the positivity of weights. Moreover, DQ rules can be constructed over non-standard geometries of the support of the nonnegative weight function and on more general polynomial spaces (e.g., hyperbolic cross polynomial space) instead of restricting to just total-order polynomial spaces. For instance, \cite{keshavarzzadeh2018numerical} considered the support of weight function to be ``U" shape while generating DQ. In fact, DQ requires relatively fewer nodes than SGQ. For example, to approximate a 10 dimensional integral with a polynomial of total order 5 as integrand, DQ requires 148 nodes while nested SGQ needs 201 nodes.\footnote{In the absence of information about functional properties of the integrand, apriori assumption on the order of the polynomial (to approximate the integrand) persists in DQ. This theoretical issue does not adversely affect the empirical advantages of DQ because one can generate DQ rules for the highest possible polynomial order for a given number of nodes, which can be restored and reused in future (See section \ref{sec:con} for a detailed discussion). Adaptive SGQ methods are capable of handling this challenge  \citep{ma2009adaptive, brumm2017using,cagnone2017adaptive, bhaduri2018efficient}. These adaptive methods are not restricted to polynomial basis functions, rather hierarchical linear or non-linear basis functions are generally used to capture the local behavior of the integrand. However, adaptive methods inherit two empirical issues -- i) These methods are generally computationally expensive; and ii) since the basis function is adaptively updated in each dimension based on properties of the integrand, quadrature rules need to be computed for each problem (i.e., cannot be reused).} 

To the best of our knowledge, the potential of moment-based numerical quadrature rules has not been explored in the econometrics literature. Thus, the contribution of the study is twofold: i) we address the bottlenecks of the traditional SGQ method by applying the recently developed DQ method \citep{keshavarzzadeh2018numerical} in maximum simulated likelihood estimation of discrete choice models; (ii) using a Monte Carlo study and an empirical application, we show the superiority of DQ over workhorse QMC methods in estimation of MMNL with varying number of random parameters (3, 5, and 10) and correlation structures (diagonal and full covariance).

The rest of the paper is organized as follows: section \ref{sec:mix} briefly describes the MMNL model and its estimation, section \ref{sec:qua} discusses univariate quadrature, multivariate quadrature, and DQ methods, section \ref{sec:mon} explains the Monte Carlo simulation design and summarizes corresponding results, section \ref{sec:emp} compares QMC methods with DQ on an empirical study, and conclusions and future work are detailed in section \ref{sec:con}. 

\newpage

\section{Mixed Multinomial Logit Model} \label{sec:mix}
Consider that the conditional indirect utility derived by decision-maker $i$ from making choice $j$ in choice situation $t$ is:
	\begin{equation}
	U_{itj} = \bm{x}^{T}_{itj} \bm{\alpha} + \bm{z}^{T}_{itj}\bm{\beta}_{i} + \varepsilon_{itj},
	\end{equation}
	where $i \in \{1,\dots,N \}$, $j \in \{1,\dots,J \}$, and $t \in \{1,\dots,T \}$.  The covariate vector $\bm{x}_{itj}$ has a fixed preference parameter vector $\bm{\alpha}$ and $\bm{z}_{itj}$ has a random, agent-specific parameter vector $\bm{\beta}_{i}$. The preference shock $\varepsilon_{itj}$ is independent across individuals, choices and time, and is identically distributed Type-I Extreme Value. Thus, the probability of choosing alternative $j$ by individual $i$ in choice situation $t$, conditional on $\bm\beta_{i}$, has a logit link: 
	\begin{equation} \label{P_itj}
	P_{itj}(\bm{\alpha},\bm{\beta}_i) = \frac{\exp\left(\bm{x}^{T}_{itj} \bm{\alpha} + \bm{z}^{T}_{itj}\bm{\beta}_{i}\right)}{\sum_{k=1}^{J} \exp\left(\bm{x}^{T}_{itk} \bm{\alpha} + \bm{z}^{T}_{itk}\bm{\beta}_{i}\right)}.
	\end{equation}
	
	For an individual $i$ who chooses alternative $j$ in choice situation $t$, we define the indicator $d_{itj} = \mathbb{I}($j$ \textrm{ chosen}|i,t )$. For the sequence of choices made by individual $i$, the conditional likelihood $\mathcal{L}_{i}(\bm{\alpha},\bm{\beta}_i)$ is: 
	\begin{equation} \label{L_i}
	\mathcal{L}_{i}(\bm{\alpha},\bm{\beta}_i) = \prod_{t=1}^{T} \prod_{j=1}^{J} [P_{itj}(\bm{\alpha},\bm{\beta}_{i})]^{d_{itj}}.
	\end{equation}
	
Consider that the random parameter $\bm{\beta}_{i}$ is multivariate normally distributed with mean $\bm{\gamma}$ and variance-covariance matrix $\bm{\Delta}$. Thus, the loglilkelihood $\ell(\bm{\psi})$ of the sample in terms of the unconditional likelihood $P_{i}(\bm{\psi})$ of individual $i$ is: 

    \begin{equation}\label{LL}
    \begin{aligned}
    & \ell(\bm{\psi}) = \sum_{i=1}^{N} \ln\Big(P_{i}(\bm{\psi})\Big) =  \sum_{i=1}^{N} \ln\Bigg( \int_{\bm{\beta}} \mathcal{L}_{i}(\bm{\alpha},\bm{\beta}) f(\bm{\beta} | \bm{\gamma}, \bm{\Delta}) d\bm{\beta} \Bigg), \\
    & \textrm{where } \bm{\psi} = \{ \bm{\alpha}, \bm{\gamma}, \bm{\Delta}\}.
    \end{aligned}
    \end{equation}

Since the sample loglikelihood $\ell(\cdot)$ in equation \ref{LL} is analytically intractable, the parameter vector $\bm{\psi}$ can be estimated by maximizing the sample's simulated loglikelihood $\widetilde{\ell}(\cdot)$:  

	\begin{equation}
	\widetilde{\ell}(\bm{\psi}) = \sum_{i=1}^{N} 
	\ln\Bigg(\sum_{r =1}^R \mathcal{L}_{i}(\bm{\alpha},\bm{\beta}_{ir})w_i(\bm{\beta}_{ir} |\bm{\gamma}, \bm{\Delta})\Bigg).
	\end{equation}

Note that $\bm{\beta}_{ir}$ and  $w_i(\bm{\beta}_{ir} |\bm{\gamma}, \bm{\Delta})$ are viewed as nodes and weights in the quadrature method, respectively. In the QMC simulation literature, nodes are generally denoted by draws and the weight $w_i(\bm{\beta}_{ir} |\bm{\gamma}, \bm{\Delta})$ attains the value of $\frac{1}{R}$ for all draws. Note that even though $\bm{\beta}_{ir}$ is a realization of $\mathcal{N}(\bm{\gamma}, \bm{\Delta})$, the model is  reparametrized in terms of the Cholesky decomposition of $\bm{\Delta}$ to ensure positive definiteness. Thus, when approximating the loglikelihood with quadrature or QMC methods, we always work with standard normal distributions.

\section{Quadrature Methods} \label{sec:qua}
\subsection{Notation}
We adopt the notation of \cite{keshavarzzadeh2018numerical} to illustrate the intuition and key results of different quadrature methods. We reconsider the integral approximation problem: 
    \begin{equation} \label{eq:approx}
        \int_\Gamma f(\bm{x}) \omega(\bm{x}) d\bm{x} \approx \sum_{q=1}^n f(\bm{x}_q) w_q,
    \end{equation}
where $\omega(\bm{x})$ is a given weight function (or a probability density function) whose support is $\Gamma \subset \mathbb{R}^d$.  A point $\bm{x} \in \mathbb{R}^d$ has components $\bm{x} = \left( x^{(1)}, x^{(2)}, \ldots, x^{(d)} \right)$. 

We define $\bm{\alpha} \in \mathbb{N}_0^d$ as a multi-index, and $\Lambda$ as a downward closed set\footnote{If $\bm{\alpha}$, $\bm{\beta} \in  \mathbb{N}_0^d$, then $\bm{\alpha} \leq \bm{\beta}$ if and only if all component-wise inequalities are true. Using this definition, a multi-index set $\Lambda$ is called \textit{downward closed} if
$\bm{\alpha} \in \Lambda \; \Longrightarrow  \;  \bm{\beta} \in \Lambda \quad  \forall \; \bm{\beta} \leq \bm{\alpha}.$} of multi-indices:

    \begin{equation*}
        \bm{\alpha} = (\alpha_1, \ldots, \alpha_d), \quad \quad \bm{x}^{\bm{\alpha}} = \prod_{j=1}^d \left( x^{(j)} \right)^{\alpha_j}, \quad \quad |\bm{\alpha}| = \sum_{j=1}^d \alpha_j.
    \end{equation*}

Our ultimate goal is to construct a set of $n$ points $\left\{ \bm{x}_q \right\}_{q=1}^n \subset \Gamma$ and positive weights $w_q > 0$ in equation \ref{eq:approx}, but we attempt to achieve this by enforcing equality in equation \ref{eq:approx} for a subspace $\Pi$ of polynomials such that:
    \begin{equation}
        \begin{aligned}\label{eq:quadrature-equality}
            & \int_\Gamma f(\bm{x}) \omega(\bm{x}) d\bm{x} = \sum_{q=1}^n f(\bm{x}_q) w_q, \quad  f \;\in\;\Pi \\
            & \Pi = \mathrm{span} \left\{ \bm{x}^ {\bm \alpha} \;\; \big| \;\; \bm \alpha \in \Lambda \right\}.
        \end{aligned} 
    \end{equation}

Thus, solving for $\left\{ \bm{x}_q \right\}_{q=1}^n$ and $w_q > 0$ using equation \ref{eq:quadrature-equality} should provide a good approximation of the integral in equation \ref{eq:approx}.

Whereas \cite{keshavarzzadeh2018numerical} proposed a numerical method to solve equation \ref{eq:quadrature-equality} for a general polynomial subspaces, we restrict discussion to \textit{total order} (represented by subscript $\mathcal{T}_{(.)}$ ) polynomial spaces with the total order being $r$: 

    \begin{equation}
        \Pi_{\mathcal{T}_r} = \mathrm{span} \left\{ \bm{x}^{\bm \alpha} \;\; \big| \;\; \bm \alpha \in \Lambda_{\mathcal{T}_r} \right\}, \quad \textrm{where }  \Lambda_{\mathcal{T}_r} = \left\{ \bm \alpha \in \mathbb{N}_0^d \;\; \big| \;\; |\bm \alpha| \leq r \right\}.
    \end{equation}

\subsection{Univariate Quadrature}
 We need to define first the basis for the polynomial space $\Pi_{\mathcal{T}_k}$. Note that a basis of orthonormal polynomials exists with elements $p_m(\cdot)$ such that $\deg p_m = m$. The family of these polynomials satisfies the following recursive relation \citep{askey1975orthogonal}:
 
    \begin{equation}\label{eq:three-term-recurrence}
        \begin{aligned}
         & x p_m(x) = \sqrt{b_m} p_{m-1}(x) + a_m p_m(x) + \sqrt{b_{m+1}} p_{m+1}(x), \\
         & a_m = (x p_m, p_m) \quad b_m = \frac{(p_m, p_m)}{(p_{m-1}, p_{m-1})}
        \end{aligned}
    \end{equation}

After characterizing the one-dimension polynomial space, we present a theorem which is the foundation of the Quadrature literature: \\
\textbf{Theorem 1 (Gaussian quadrature)}
Let $x_{1},\ldots,x_{n}$ be the roots of the $n^{th}$ orthogonal polynomial $p_n(x)$ and let $w_{1},\ldots,w_{n}$ be the solution of the system of equations
\begin{equation}\label{THM1_0}
\sum_{q=1}^n p_j(x_q) w_q =
\begin{cases}
  \sqrt{b_0}, & \textrm{if } j=0\\
  0, & \textrm{for } j=1,\ldots,n-1.\\
\end{cases}
\end{equation}
Then $x_q \in \Gamma$ and $w_q>0$ for $q=1,2,\ldots,n$, and
\begin{equation}\label{THM1_1}
\displaystyle \int_{\Gamma}  p(x) \omega(x) dx = \sum_{q=1}^n p(x_q) w_q
\end{equation}
holds for all polynomials $p \in \Pi_{\mathcal{T}_{2n-1}}$.\\

According to Theorem 1,  nodes and weights in equation \ref{THM1_1} (which is a one-dimensional version of equation \ref{eq:quadrature-equality}) can be exactly obtained by solving the system of equations (moment-matching conditions) summarized in equation \ref{THM1_0}. A more intuitive implication of Theorem 1 is that if the integrand can be exactly specified on a polynomial space of order $2n-1$, only $n$ nodes are required to compute the corresponding univariate integral precisely. \cite{golub1969calculation} and \cite{davis2007methods} provide a detailed procedure to compute this univariate quadrature rule.    

\subsection{Multivariate Quadrature}
In product rules, univariate quadrature can be simply extended to multivariate quadrature using a tensor product. More specifically, the weight function $\omega(\bm{x})$ and its support $\Gamma$ can be written as follows: 
\begin{equation*}
  \Gamma = \times_{j=1}^d \Gamma_j, \quad \quad \quad \quad \quad \omega(\bm{x}) = \prod_{j=1}^d \omega_j\left(x^{(j)}\right),
\end{equation*}
where $\Gamma_j \subset \mathbb{R}$  is univariate domain and $\omega_j(\cdot)$ is a univariate weight. If $p^{(j)}_n(\cdot)$ is the univariate orthonormal polynomial family corresponding to $\omega_j$ over $\Gamma_j$, then the family of multivariate polynomials orthonormal under $\omega$ can be written as: 
\begin{equation*}
  \pi_{\bm{\alpha}}(\bm{x}) = \prod_{j=1}^d p^{(j)}_{\alpha_j}\left(x^{(j)}\right), \quad \quad \quad \quad \quad \bm{\alpha} \in \mathbb{N}_0^d.
\end{equation*}

The corresponding polynomial space is: $\Pi_{\mathcal{T}_r} = \mathrm{span} \left\{ \pi_{\bm{\alpha}} \;\; \big| \;\; \bm{\alpha} \in \Lambda_{\mathcal{T}_r} \right\}$. After characterizing the polynomial subspace, the moment-matching conditions of Theorem 1 can be extended to the multivariate case as follows:\\

\textbf{Proposition 1}
 Let $\Lambda$ be a multi-index set with $\bm{0} \in \Lambda$. Suppose that $\bm x_{1},\ldots,\bm x_{n}$ and $w_{1},\ldots,w_{n}$ are the solution of the system of equations
\begin{equation}\label{PRO2_0}
\sum_{q=1}^n \pi_{\bm \alpha}(\bm x_{q}) w_{q} =
\begin{cases}
  1/\pi_{\bm{0}}, &\textrm{if } \bm \alpha= \bm 0\\
  0, &\textrm{if } \bm \alpha \in \Lambda\backslash\{\bm 0\}\\
\end{cases}
\end{equation}
then
\begin{equation}\label{PRO2_1}
\displaystyle \int_{\bm \Gamma} \omega(\bm x) \pi(\bm x) d\bm x = \sum_{q=1}^n \pi(\bm x_{q}) w_{q}
\end{equation}
holds for all polynomials $\pi \in \mathrm{\Pi}_{\Lambda}$.

Note that unlike Theorem 1, the above proposition neither guarantees the positivity of weights nor ensures that nodes belong to support $\Gamma$. Although sparse grid quadrature (SGQ) provides an efficient way to combine multiple dimensions so as to reduce the function evaluations, it does not provide remedy for these issues. We did not consider SGQ in this study, because: a) in our initial test runs, negative weights in SGQ led to complex (imaginary) loglikelihood values in estimation of MMNL with full variance-covariance matrix; b) based on extensive simulations studies, \cite{keshavarzzadeh2018numerical} confirmed that designed quadrature (DQ) requires many fewer nodes than SGQ.  \cite{heiss2008likelihood} can be referred for intuitive and theoretical discussion on SGQ rules.

\subsection{Designed Quadrature (DQ)}
DQ solves a relaxed version ofthe moment-matching conditions given in equation \ref{PRO2_0}, which enforces positivity of weights and also ensure nodes to fall in the support of the probability density function. \cite{keshavarzzadeh2018numerical} reformulates the moment-matching conditions as follows: 

For a given index set $\Lambda$ with size $M = |\Lambda|$, consider the matrix $\bm{X} \in \mathbb{R}^{d \times n}$ with columns $\bm{x}_j$, and let $\bm{w} \in \mathbb{R}^n$ be a vector containing the $n$ weights. Let $\bm{V}(\bm{X}) \in \mathbb{R}^{M \times n}$ denote the Vandermonde-like matrix with entries
    \begin{equation}\label{eq:Vandermonde-def}
        \left(V \right)_{k,j} = \pi_{\bm{\alpha}(k)}\left(\bm{x}_j\right), \quad k=1, \ldots, M \quad j = 1, \ldots, n ,
    \end{equation}
where elements of $\Lambda$ are considered with ordering $\bm{\alpha}(1), \ldots \bm{\alpha}(M)$ and $\bm{\alpha}(1) = \bm{0}$. The system \eqref{PRO2_0} can then be written as: 
    \begin{equation}\label{eq:MMTRUE}
        \bm{V}\left(\bm{X}\right) \bm{w} = \bm{e}_1/\pi_{\bm{0}},
    \end{equation}
where $\bm{e}_1 = (1, 0, 0, \ldots, 0)^T \in \mathbb{R}^M$. Instead of solving the moment-matching conditions exactly in equation \ref{eq:MMTRUE}, \cite{keshavarzzadeh2018numerical} proposed to obtain the approximate solution $\left(\bm{X},\bm{w}\right)$ that satisfies: 
    \begin{equation}\label{eq:epsilon}
        \left\| \bm{V}\left(\bm{X}\right) \bm{w} - \bm{e}_1/\pi_{\bm{0}} \right\|_2 = \epsilon \geq 0.
    \end{equation}

In fact, \cite{keshavarzzadeh2018numerical} provide bounds on the integral error \\
$ \left| \int f(\bm{x}) \omega(\bm{x}) d{\bm{x}} - \sum_{q=1}^n  f(\bm{x}_q)w_q \right| $ in terms of tolerance $\epsilon$, which is computable for a given quadrature rule.  

Thus, for a given polynomial subspace, DQ aims to compute nodes $\bm{X} = \left\{ \bm{x}_1, \ldots, \bm{x}_n \right\}\in \Gamma^n$ and positive weights $\bm{w} \in (0, \infty)^n$ that solves the following constrained optimization problem:
\begin{equation}\label{eq:opt}
\begin{array}{r l l}
  \displaystyle \mathop{\min}_{\bm X, \bm w} & \displaystyle ||\bm V(\bm X) \bm w - \bm{e}_1/\pi_{\bm{0}}||_2  &  \\
  \text{subject to} & \bm x_j \in \Gamma, & j=1, \ldots, n \\
                    & w_j> \bm 0, & j =1, \ldots, n.
\end{array}
\end{equation}

Readers can refer to \cite{keshavarzzadeh2018numerical} for more insights about strategies (e.g., constrained optimization problem) to solve the above optimization problem.

\subsection{Discussion} 
In the context of this study, we explore possibilities of approximating the unconditional choice probability integral (see equation \ref{LL}) in MMNL using DQ. We assume that the conditional choice probability (integrand in equation \ref{LL}) can be approximated on total order polynomial space. Since properties of the integrand vary with the data generating process and are thus not known beforehand, the performance of the approximation will depend on the assumed order of the polynomial space. Moreover, whereas SGQ predetermines the exact number of nodes based on the order ($r$) of the polynomial space and dimension of the integral, DQ rules can be obtained (i.e., the optimization problem in equation \ref{eq:opt} can be solved) for various possible number of nodes ($n$). Thus, for a given integral dimension, one can generate DQ rules for different total order ($r$) polynomial spaces and different number of nodes ($n$). 

In both parametric and non-parametric MMNL models, the choice probability integral can generally be reparameterized such that the weight function $\omega(.)$ in DQ turns out to be a probability density function of a standard normal  (e.g., normal or lognormal mixing distributions) or a standard uniform distribution (e.g., semi-parametric logit-mixed logit model)\footnote{The support $\Gamma$ of standard normal and standard distributions are whole real line and $[0,1]$, respectively}. Just as in QMC methods we can generate, store and reuse DQ rules for both standardized distributions. Thus, DQ offers the same flexibility. The researcher can solve the optimization problem in equation \ref{eq:opt} beforehand for different combinations of dimensions, order of polynomial, and number of nodes, and then reuse the stored nodes and weights. 

In sum, DQ may appear more cumbersome than QMC methods at first, but re-usability of the nodes and weights not only makes DQ equally easy to implement in practice and in fact, even fewer function evaluations are needed (i.e., lower computation time is achieved). Nevertheless, for a given dimension of integral, whereas QMC needs tuning of the number of draws to get stable parameter estimates, DQ requires to tune the total order of polynomial spaces and the corresponding number of draws. In the next section we conduct a detailed simulation study to make recommendations about selection of these parameters in the context of MMNL.

\section{Monte Carlo Study} \label{sec:mon}
\subsection{Simulation Design}
The objective of the simulation study is to evaluate the performance of DQ relative to QMC methods in MMNL estimation. We considered modified latin hypercube sampling (MLHS) \citep{hess2006use} and randomized and scrambled Halton sequences \citep{bhat2003simulation} as representative QMC methods. In the data generating process (DGP), we considered a sample of 1000 decision makers who are assumed to choose a utility maximizing alternative from a set of five alternatives across five choice situations. Since the number of random parameters governs the dimension of the choice probability integral in MMNL, we compared performance of DQ and QMC methods in MMNL with three, five, and ten normally-distributed random parameters. For each random parameter scenario, we considered two covariance structures: zero (diagonal) and full covariance across random parameters. The considered covariance matrix $\bm{\Delta}_{full \; cov.}^{5}$ for five random parameters is illustrated below; similar structures were considered for dimensions three and ten. This sensitivity analysis is crucial because the performance of DQ depends on the smoothness of integrand (i.e., conditional likelihood), which  in turn depends on the structure of the covariance matrix. 

\[
\bm{\Delta}_{full \; cov.}^{5} =
  \begin{bmatrix}
    1.5 & 0.5 & 0.5 & 0.5 & 0.5  \\
    0.5 & 1 & 0.5 & 0.5 & 0.5  \\
    0.5 & 0.5 & 1 & 0.5 & 0.5  \\
    0.5 & 0.5 & 0.5 & 1 & 0.5  \\
    0.5 & 0.5 & 0.5 & 0.5 & 1.5  \\
  \end{bmatrix} 
\]

We generated  150 datasets in total for each covariance structure: 50 datasets for each random parameter scenario. For each of 300 datasets, we performed maximum simulated likelihood estimation (with analytical gradient) using a different number of QMC draws and different total order polynomial subspaces and nodes of DQ. We summarize results by computing the following five metrics across resamples: i) average \textit{loglikelihood} at convergence, ii) absolute percentage bias (\textit{APB}), iii) average estimation time, iv) average number of loglikelihood evaluations in the estimation, and v) the \textit{t-distributed test statistic} under the null hypothesis that the point estimate is equal to the true population parameter. As the test statistic gets smaller, we become more confident that the estimated parameter is close to the population parameter. 

We compute the APB and the \textit{t-distributed test statistic} of a parameter for a sample  as follows: 

\begin{equation*}
\textrm{APB} = \left|\dfrac{\textrm{Parameter Estimate - True Parameter Value}}{\textrm{True Parameter Value}}\right| \times 100,  
\end{equation*}

\begin{equation*}
   \textrm{Test statistic} =  \dfrac{\textrm{Mean of the Point Estimate across Resamples - True Parameter Value}}{\textrm{Finite sample standard error (FSSE)}}.
\end{equation*}

The mean values of APB and the test statistic across all parameters and resamples are reported for succinctness. To avoid empirical identification issues, we computed FSSE, APB, and t-value for  parameter ratios. We wrote MATLAB code to generate DQ rules and perform MMNL estimation. DQ rules were generated beforehand, stored, and reused for estimation. While generating DQ rules, we considered tolerance $\epsilon$ in equation \ref{eq:epsilon} to be $10^{-8}$. We performed sensitivity analysis with tighter tolerances but those did not improve accuracy.     

\subsection{Results and Discussion}
The results of the Monte Carlo study for random parameters (integral dimensions) three, five, and ten are summarized in Tables \ref{tab:D3}, \ref{tab:D5}, and \ref{tab:D10}, respectively.

As the dimension of the integral increases, the minimum number of nodes required to generate the appropriate DQ rule at a given polynomial order ($r$, also known as accuracy level) increases. For example, we could generate the DQ rule for higher-order $r=6$ with just 30 nodes for three dimensions (see Table \ref{tab:D3}), but to solve the DQ optimization problem (up to a prespecified tolerance $\epsilon$) for the same order in five dimensions needed at least 100 nodes (see Table \ref{tab:D5}). Also, for a given dimension of the integral, more nodes are required to generate DQ in higher-order polynomial spaces. For example, we could generate the DQ rule for ten dimensions with 100 nodes for a polynomial of order $r=4$, but needed a minimum of 200 nodes for $r=5$ (see Table \ref{tab:D10}).

We now compare the model fit (loglikelihood) of DQ and QMC methods. In the diagonal variance-covariance case, DQ outperformed both QMC methods by a significant margin, even when the DQ rule was generated on polynomial spaces with relatively low order. For the five-dimensional case, DQ achieved a similar model fit (loglikelihood: -5354.6) with just 100 nodes at $r=6$ as of the fit obtained using 1000 Halton and MLHS draws (loglikelihood: -5353.8 and -5353.9). In fact, DQ with 100 nodes at $r=4$ (loglikelihood: -5017.7) outperformed Halton and MLHS with 300 draws (loglikelihood: -5021.6 and -5020.9) in approximating the higher (i.e., ten) dimensional integral. The model fit values of Halton and MLHS are indistinguishable.

DQ also outperformed QMC methods in the full variance-covariance scenario, but a higher order of polynomial subspaces are desirable in this non-independent case. These observations are aligned with intuition: introducing covariance makes the integrand more complex \citep{abay2015evaluating}, which can be better approximated on higher-order polynomial subspaces. For example, in the case of five random parameters with a full covariance DGP, whereas DQ could achieve a model fit of -5759.5 with 200 nodes at $r=7$, MLHS and Halton required 300 draws to achieve virtually the same model fit; however, 300 nodes of DQ at $r=5$ (loglikelihood: -5061.0) were slightly outperformed by 300 Halton and MLHS draws (loglikelihood: -5758.4 and -5758.6). As expected, we generally observed that increasing the order of polynomial subspaces results in a better model fit. As a general trend, across all dimensions and covariance structures, the highest order in DQ (r = 7, 7, and 5 for dimensions 3, 5, and 10) resulted in better model fit than those of QMC methods at a given number of draws.

Across DQ and QMC methods, parameter recovery metrics -- 
APB -- decreases with an increase in the number of draws (or nodes). Consistent with the model fit, DQ surpassed MLHS by a significant margin in recovering true parameters if the variance-covariance matrix is diagonal. For instance, DQ could achieve similar APB value with 200 nodes (on polynomial space of order $r=6$) as of what we obtained using QMC with 1000 draws in the DGP with five random parameters (see Table \ref{tab:D5}). In fact, DQ also recovered parameters better than QMC methods across correlated covariance structures, but at higher-order polynomial subspaces. For example, in DGP with five highly correlated random parameters, APB using 200 QMC draws is 15.8\%, but for the same number of DQ nodes whereas APB is relatively higher  at $r=5$ (16.8\%), it is relatively lower at $r=7$  (13.9\%) (see Table \ref{tab:D5}). We derived similar insights by comparing t-distributed test statistic of DQ and QMC methods across the considered DGP scenario.  

One may speculate that DQ might take more evaluations of the loglikelihood than those required in QMC to achieve convergence, leading to higher estimation time than QMC for the same number of draws/nodes. To test this hypothesis, we present the average estimation time, and loglikelihood function evaluation across all resamples. The results indicate that DQ requires the same or fewer loglikelihood evaluations than those needed in QMC methods. Since each loglikelihood evaluation takes the same time in both DQ and QMC for a given number of draws, the estimation time of DQ is similar or lower than that of QMC methods. Thus, the superiority of DQ in terms of model fit and parameter recovery directly translates into computational time savings. 

In sum, for a given number of draws/nodes, better model fit, and more precise parameter recovery in DQ across all dimensions and covariance structures make it a computationally-efficient substitute to QMC methods in practice.

\begin{landscape}
\vspace*{\fill}

\begin{table}[H]
		\centering
		\caption{\footnotesize{Comparison of designed quadrature and quasi Monte Carlo sequences (Monte Carlo study, random parameters=3)}}
		\label{tab:D3}
		\resizebox{1.6\textwidth}{!}{
\begin{tabular}{ccccccccccccccccccccccccc}
\hline
\textbf{} & \multicolumn{4}{c}{\textbf{(-)Loglikelihood}} & \textbf{} & \multicolumn{4}{c}{\textbf{Absolute percentage bias}} & \textbf{} & \multicolumn{4}{c}{\textbf{Estimation time (seconds)}} & \textbf{} & \multicolumn{4}{c}{\textbf{Loglikelihood function counts}} & \textbf{} & \multicolumn{4}{c}{\textbf{t-value}} \\ \cline{1-5} \cline{7-10} \cline{12-15} \cline{17-20} \cline{22-25} 
\textbf{Draws} & \textbf{Halton} & \textbf{MLHS} & \multicolumn{2}{c}{\textbf{DQ}} & \textbf{} & \textbf{Halton} & \textbf{MLHS} & \multicolumn{2}{c}{\textbf{DQ}} & \textbf{} & \textbf{Halton} & \textbf{MLHS} & \multicolumn{2}{c}{\textbf{DQ}} & \textbf{} & \textbf{Halton} & \textbf{MLHS} & \multicolumn{2}{c}{\textbf{DQ}} & \textbf{} & \textbf{Halton} & \textbf{MLHS} & \multicolumn{2}{c}{\textbf{DQ}} \\ \cline{1-5} \cline{7-10} \cline{12-15} \cline{17-20} \cline{22-25} 
\textbf{} & \textbf{} & \textbf{} & \textbf{r=6} & \textbf{r=7} & \textbf{} & \textbf{} & \textbf{} & \textbf{r=6} & \textbf{r=7} & \textbf{} & \textbf{} & \textbf{} & \textbf{r=6} & \textbf{r=7} & \textbf{} & \textbf{} & \textbf{} & \textbf{r=6} & \textbf{r=7} & \textbf{} & \textbf{} & \textbf{} & \textbf{r=6} & \textbf{r=7} \\ \cline{1-5} \cline{7-10} \cline{12-15} \cline{17-20} \cline{22-25} 
\textbf{Diagonal} &  &  &  &  &  &  &  &  &  &  &  &  &  &  &  &  &  &  &  &  &  &  &  &  \\ 
30 & 5752.9 & 5749.8 & 5725.3 &  &  & 7.1 & 7.7 & 5.2 &  &  & 2.4 & 2.3 & 1.9 &  &  & 24 & 24 & 23 &  &  & 0.31 & 0.28 & 0.37 &  \\
50 & 5739.7 & 5740.7 & 5725.1 & 5724.5 &  & 6.4 & 6.9 & 5.2 & 5.2 &  & 3.8 & 3.7 & 3.4 & 3.4 &  & 24 & 24 & 23 & 23 &  & 0.22 & 0.24 & 0.32 & 0.34 \\
100 & 5732.7 & 5731.4 & 5724.8 & 5724.4 &  & 5.4 & 6.1 & 5.1 & 5.2 &  & 7.3 & 7.2 & 6.9 & 6.7 &  & 23 & 24 & 24 & 23 &  & 0.32 & 0.22 & 0.33 & 0.33 \\
1000 & 5724.5 & 5724.6 &  &  &  & 5.2 & 5.2 &  &  &  & 73.3 & 73.9 &  &  &  & 23 & 23 &  &  &  & 0.34 & 0.33 &  &  \\ \cline{1-5} \cline{7-10} \cline{12-15} \cline{17-20} \cline{22-25} 
\textbf{Full covariance} &  &  &  &  &  &  &  &  &  &  &  &  &  &  &  &  &  &  &  &  &  &  &  &  \\ \cline{1-5} \cline{7-10} \cline{12-15} \cline{17-20} \cline{22-25} 
30 & 5921.0 & 5916.0 & 5860.4 &  &  & 15.1 & 14.5 & 13.0 &  &  & 2.2 & 2.0 & 1.6 &  &  & 22 & 21 & 20 &  &  & 0.53 & 0.42 & 0.26 &  \\
50 & 5865.4 & 5863.2 & 5849.0 & 5835.0 &  & 14.1 & 12.6 & 11.5 & 11.4 &  & 3.5 & 3.5 & 3.0 & 2.9 &  & 23 & 23 & 21 & 21 &  & 0.37 & 0.37 & 0.36 & 0.19 \\
100 & 5827.5 & 5827.1 & 5827.9 & 5810.7 &  & 10.8 & 10.5 & 10.7 & 9.3 &  & 7.0 & 7.1 & 6.4 & 6.4 &  & 24 & 24 & 23 & 23 &  & 0.31 & 0.20 & 0.28 & 0.22 \\
1000 & 5794.5 & 5794.7 &  &  &  & 8.7 & 9.2 &  &  &  & 73.0 & 73.0 &  &  &  & 25 & 24 &  &  &  & 0.21 & 0.18 &  &  \\
\hline
\end{tabular}		
}
\end{table}	

\vspace*{\fill}
\end{landscape}

\begin{landscape}
\vspace*{\fill}

\begin{table}[H]
		\centering
		\caption{\footnotesize{Comparison of designed quadrature (DQ) and quasi Monte Carlo (QMC) sequences (Monte Carlo study, random parameters=5)}}
		\label{tab:D5}
		\resizebox{1.6\textwidth}{!}{
\begin{tabular}{cccccccccccccccccccccccccccccc}
\hline 
\textbf{} & \multicolumn{5}{c}{\textbf{(-)Loglikelihood}} & \textbf{} & \multicolumn{5}{c}{\textbf{Absolute percentage bias}} & \textbf{} & \multicolumn{5}{c}{\textbf{Estimation time (seconds)}} & \textbf{} & \multicolumn{5}{c}{\textbf{Loglikelihood function counts}} & \textbf{} & \multicolumn{5}{c}{\textbf{t-value}} \\ \cline{1-6} \cline{8-12} \cline{14-18} \cline{20-24} \cline{26-30} 
\textbf{Draws} & \textbf{Halton} & \textbf{MLHS} & \multicolumn{3}{c}{\textbf{DQ}} & \textbf{} & \textbf{Halton} & \textbf{MLHS} & \multicolumn{3}{c}{\textbf{DQ}} & \textbf{} & \textbf{Halton} & \textbf{MLHS} & \multicolumn{3}{c}{\textbf{DQ}} & \textbf{} & \textbf{Halton} & \textbf{MLHS} & \multicolumn{3}{c}{\textbf{DQ}} & \textbf{} & \textbf{Halton} & \textbf{MLHS} & \multicolumn{3}{c}{\textbf{DQ}} \\ \cline{1-6} \cline{8-12} \cline{14-18} \cline{20-24} \cline{26-30} 
\textbf{} & \textbf{} & \textbf{} & \textbf{r=5} & \textbf{r=6} & \textbf{r=7} & \textbf{} & \textbf{} & \textbf{} & \textbf{r=5} & \textbf{r=6} & \textbf{r=7} & \textbf{} & \textbf{} & \textbf{} & \textbf{r=5} & \textbf{r=6} & \textbf{r=7} & \textbf{} & \textbf{} & \textbf{} & \textbf{r=5} & \textbf{r=6} & \textbf{r=7} & \textbf{} & \textbf{} & \textbf{} & \textbf{r=5} & \textbf{r=6} & \textbf{r=7} \\ \cline{1-6} \cline{8-12} \cline{14-18} \cline{20-24} \cline{26-30} 
\multicolumn{1}{c}{\textbf{Diagonal}} \\ \cline{1-6} \cline{8-12} \cline{14-18} \cline{20-24} \cline{26-30} 
50 & 5390.8 & 5388.1 & 5359.4 &  &  &  & 8.8 & 9.5 & 6.1 &  &  &  & 11 & 11 & 10 &  &  &  & 33 & 34 & 30 &  &  &  & 0.36 & 0.28 & 0.38 &  &  \\
100 & 5371.7 & 5371.8 & 5355.5 & 5354.6 &  &  & 7.2 & 6.8 & 5.6 & 5.5 &  &  & 20 & 20 & 20 & 19 &  &  & 31 & 31 & 30 & 30 &  &  & 0.28 & 0.33 & 0.28 & 0.28 &  \\
200 & 5363.3 & 5362.5 & 5354.6 & 5353.2 & 5352.8 &  & 6.2 & 5.8 & 5.5 & 5.3 & 5.4 &  & 41 & 39 & 39 & 38 & 38 &  & 32 & 30 & 30 & 30 & 29 &  & 0.20 & 0.26 & 0.26 & 0.23 & 0.23 \\
300 & 5359.0 & 5359.1 & 5354.0 & 5352.2 & 5352.3 &  & 5.9 & 5.9 & 5.5 & 5.3 & 5.2 &  & 60 & 60 & 57 & 57 & 57 &  & 30 & 30 & 29 & 30 & 30 &  & 0.23 & 0.24 & 0.24 & 0.23 & 0.21 \\
1000 & 5353.8 & 5353.9 &  &  &  &  & 5.2 & 5.3 &  &  &  &  & 196 & 195 &  &  &  &  & 30 & 29 &  &  &  &  & 0.21 & 0.21 &  &  &  \\ \cline{1-6} \cline{8-12} \cline{14-18} \cline{20-24} \cline{26-30} 
\multicolumn{1}{c}{\textbf{Full covariance}} \\ \cline{1-6} \cline{8-12} \cline{14-18} \cline{20-24} \cline{26-30} 
50 & 5895.5 & 5886.2 & 5840.3 &  &  &  & 23.4 & 24.7 & 18.9 &  &  &  & 12 & 11 & 10 &  &  &  & 30 & 28 & 25 &  &  &  & 0.33 & 0.36 & 0.57 &  &  \\
100 & 5817.8 & 5819.2 & 5785.6 & 5794.0 &  &  & 18.3 & 18.5 & 17.8 & 16.9 &  &  & 23 & 23 & 22 & 21 &  &  & 30 & 30 & 29 & 28 &  &  & 0.21 & 0.24 & 0.32 & 0.29 &  \\
200 & 5773.7 & 5772.5 & 5772.5 & 5761.3 & 5759.5 &  & 15.9 & 15.8 & 16.8 & 15.0 & 13.9 &  & 46 & 45 & 44 & 43 & 44 &  & 30 & 30 & 29 & 28 & 29 &  & 0.16 & 0.14 & 0.26 & 0.16 & 0.24 \\
300 & 5758.4 & 5758.6 & 5761.0 & 5749.5 & 5743.5 &  & 14.8 & 14.9 & 15.8 & 14.2 & 12.9 &  & 72 & 71 & 68 & 65 & 66 &  & 31 & 31 & 30 & 29 & 29 &  & 0.15 & 0.13 & 0.21 & 0.17 & 0.16 \\
1000 & 5733.1 & 5733.0 &  &  &  &  & 12.5 & 12.7 &  &  &  &  & 238 & 231 &  &  &  &  & 31 & 30 &  &  &  &  & 0.13 & 0.11 &  &  &  \\ 
\hline
\end{tabular}}
\end{table}

\begin{table}[H]
		\centering
		\caption{\footnotesize{Comparison of designed quadrature (DQ) and quasi Monte Carlo (QMC) sequences (Monte Carlo study, random parameters=10)}}
		\label{tab:D10}
		\resizebox{1.6\textwidth}{!}{
\begin{tabular}{cccccccccccccccccccccccccccccc}
\hline
\textbf{} & \multicolumn{5}{c}{\textbf{(-)Loglikelihood}} & \textbf{} & \multicolumn{5}{c}{\textbf{Absolute percentage bias}} & \textbf{} & \multicolumn{5}{c}{\textbf{Estimation time (seconds)}} & \textbf{} & \multicolumn{5}{c}{\textbf{Loglikelihood function counts}} & \textbf{} & \multicolumn{5}{c}{\textbf{t-value}} \\ \cline{1-6} \cline{8-12} \cline{14-18} \cline{20-24} \cline{26-30} 
\textbf{Draws} & \textbf{Halton} & \textbf{MLHS} & \multicolumn{3}{c}{\textbf{DQ}} & \textbf{} & \textbf{Halton} & \textbf{MLHS} & \multicolumn{3}{c}{\textbf{DQ}} & \textbf{} & \textbf{Halton} & \textbf{MLHS} & \multicolumn{3}{c}{\textbf{DQ}} & \textbf{} & \textbf{Halton} & \textbf{MLHS} & \multicolumn{3}{c}{\textbf{DQ}} & \textbf{} & \textbf{Halton} & \textbf{MLHS} & \multicolumn{3}{c}{\textbf{DQ}} \\ \cline{1-6} \cline{8-12} \cline{14-18} \cline{20-24} \cline{26-30} 
\textbf{} & \textbf{} & \textbf{} & \textbf{r=3} & \textbf{r=4} & \textbf{r=5} & \textbf{} & \textbf{} & \textbf{} & \textbf{r=3} & \textbf{r=4} & \textbf{r=5} & \textbf{} & \textbf{} & \textbf{} & \textbf{r=3} & \textbf{r=4} & \textbf{r=5} & \textbf{} & \textbf{} & \textbf{} & \textbf{r=3} & \textbf{r=4} & \textbf{r=5} & \textbf{} & \textbf{} & \textbf{} & \textbf{r=3} & \textbf{r=4} & \textbf{r=5} \\ \cline{1-6} \cline{8-12} \cline{14-18} \cline{20-24} \cline{26-30} 
\textbf{Diagonal} & \textbf{} & \textbf{} & \textbf{} & \textbf{} & \textbf{} & \textbf{} & \textbf{} & \textbf{} & \textbf{} & \textbf{} & \textbf{} & \textbf{} & \textbf{} & \textbf{} & \textbf{} & \textbf{} & \textbf{} & \textbf{} & \textbf{} & \textbf{} & \textbf{} & \textbf{} & \textbf{} & \textbf{} & \textbf{} & \textbf{} & \textbf{} & \textbf{} & \textbf{} \\
100 & 5035.3 & 5034.3 & 5022.5 & 5017.7 &  &  & 23.5 & 23.2 & 17.7 & 14.4 &  &  & 248 & 237 & 225 & 226 &  &  & 99 & 89 & 84 & 83 &  &  & 0.58 & 0.53 & 0.50 & 0.45 &  \\
200 & 5026.1 & 5024.7 & 5019.6 & 5015.1 & 5015.5 &  & 17.7 & 17.3 & 14.9 & 12.5 & 11.8 &  & 451 & 448 & 437 & 421 & 422 &  & 84 & 85 & 81 & 75 & 76 &  & 0.46 & 0.40 & 0.42 & 0.38 & 0.36 \\
300 & 5021.6 & 5020.9 & 5018.4 & 5014.7 &  &  & 15.8 & 16.2 & 14.7 & 10.9 &  &  & 667 & 669 & 673 & 605 &  &  & 83 & 85 & 84 & 71 &  &  & 0.38 & 0.39 & 0.43 & 0.30 &  \\ \cline{1-6} \cline{8-12} \cline{14-18} \cline{20-24} \cline{26-30} 
\textbf{Full covariance} &  &  &  &  &  &  &  &  &  &  &  &  &  &  &  &  &  &  &  &  &  &  &  &  &  &  &  &  &  \\ \cline{1-6} \cline{8-12} \cline{14-18} \cline{20-24} \cline{26-30} 
100 & 5914.1 & 5912.3 & 5895.7 & 5881.4 &  &  & 85.9 & 85.9 & 80.1 & 75.8 &  &  & 236 & 235 & 230 & 228 &  &  & 87 & 86 & 87 & 84 &  &  & 0.36 & 0.36 & 0.37 & 0.30 &  \\
200 & 5875.6 & 5873.1 & 5867.5 & 5858.7 & 5859.0 &  & 70.8 & 74.3 & 69.1 & 69.1 & 65.9 &  & 474 & 486 & 431 & 443 & 430 &  & 87 & 91 & 76 & 80 & 77 &  & 0.28 & 0.28 & 0.26 & 0.25 & 0.25 \\
300 & 5857.1 & 5856.2 & 5863.0 & 5847.1 &  &  & 66.7 & 64.0 & 63.0 & 54.6 &  &  & 702 & 684 & 661 & 659 &  &  & 87 & 81 & 80 & 78 &  &  & 0.24 & 0.22 & 0.28 & 0.20 & \\
\hline
\end{tabular} 
}
\end{table}	

\vspace*{\fill}
\end{landscape}

\section{Empirical Study} \label{sec:emp}
We now compare the performance of DQ and QMC while studying the preference of travelers in New York City (NYC) for mobility-on-demand (MoD) services (e.g., Uber and UberPool).  

\subsection{Experiment Design}
We conducted a stated preference survey in NYC. The survey included a discrete choice experiment (DCE) in which each respondent was asked to choose the best and the worst travel mode from a set of three choices: Uber (without ridesharing), UberPool (with ridesharing), and their current travel mode (the one used most often on their most frequent trips). We first conducted a pilot study (N=298) using D-efficient design with zero priors in February 2017. We then used prior parameter estimates from the pilot study to create a pivot-efficient design\footnote{In pivot-efficient designs, attribute levels shown to the respondents are pivoted from reference alternatives for each respondent. In this study, the travel mode used on the most frequent trips was considered as the reference alternative.} with 6 blocks (7 choice situations per block). Table \ref{tab:attribute} shows the attribute levels of the DCE design and an instance of choice situation. More details about the experiment design can be found in \cite{liu2018framework}. We conducted the main study during October-November 2017. Preferences of 1000 respondents were used in estimation.

\begin{table}[h!]
		\centering
		\caption{\footnotesize{Experiment Design for Mode Choice Study}}
		\label{tab:attribute}
		\begin{adjustwidth}{0cm}{}
		\resizebox{1\textwidth}{!}{
		\begin{tabular}{lccc}
\hline
\multicolumn{4}{c} {\textbf{Attribute Levels in the Experiment Design}} \\ \hline
 & \textbf{Uber (Without Ridesharing)} & \textbf{UberPool (With Ridesharing)} & \textbf{Current Mode} \\ \hline
Walking and Waiting Time & 25\%, 50\%, 75\%, 100\% & 25\%, 50\%, 75\%, 100\% & asked (100\%) \\
In-vehicle Travel Time & 80\%, 95\%, 110\%, 125\% & 90\%, 105\%, 120\%,135\% & asked (100\%) \\
\begin{tabular}[c]{@{}l@{}}Trip Cost Per Mile (\$)\\ (Excluding Parking Cost)\end{tabular} & \begin{tabular}[c]{@{}c@{}}$0.55$, $0.70$, $0.85$, $1.0$, $1.2$\\ \end{tabular} & \begin{tabular}[c]{@{}c@{}}$0.45$, $0.60$, $0.70$, $0.80$\\ \end{tabular} & asked or computed \\
Parking Cost & 0 & 0 & asked \\
Powertrain & Gas, Electric & Gas, Electric & Gas \\
Automation & Yes, No & Yes, No & No \\ \hline

\multicolumn{4}{c} {\textbf{Instance of a Choice Situation}} \\ \hline
 & \textbf{Uber (Without Ridesharing)} & \textbf{UberPool (With Ridesharing)} & \textbf{Current Mode: Car} \\ \hline
Walking and Waiting time & 6 minutes & 9 minutes & 12 minutes \\
In-vehicle Travel Time & 38 minutes & 50 minutes & 48 minutes \\
\begin{tabular}[c]{@{}l@{}}Trip Cost\\ (Excluding Parking Cost)\end{tabular} & \$11 & \$8 & \$6 \\
Parking Cost & -- & -- & \$6 \\
Powertrain & Electric & Gas & Gas \\
Automation & Service with Driver & Automated (No Driver) & -- \\ \hline
\multicolumn{4}{l} {\textbf{Note:} All \% are relative to the reference alternative. } \\
\end{tabular}}
\end{adjustwidth}
\end{table}	

\subsection{Estimation and Results} 
We considered marginal utilities of all five alternative-specific variables to be normally-distributed. This specification led to a five dimensional integral in MMNL estimation. Similar to the simulation study, we compared DQ against randomized and scrambled Halton draws abd MLHS in this empirical study. The number of draws/nodes was varied from 100 to 1000. We considered 50 different starting values and for each starting value, 13 models were estimated considering different QMC draws and DQ nodes. 

Table \ref{tab:LIKCASE} summarizes the average of model fit, estimation time and loglikelihood function evaluations across different starting values. Across all considered scenarios, MLHS resulted in similar model fit as obtained using Halton draws. The performance of DQ is consistent with the Monte Carlo study -- whereas QMC methods dominated DQ when rules were generated at the lower order $r=5$, DQ generated at the higher order $r=6$ always outperformed QMC methods across all considered draws. In fact, 200 nodes in DQ at order $r=6$ could achieve better model fit (loglikelihood: -5445.2) than those of 1000 Halton (loglikelihood: -5453.0) or MLHS (loglikelihood: -5453.3) draws. These gains directly translate into computational efficiency as DQ and QMC methods take similar number of loglikelihood function evaluations to achieve convergence.

\begin{table}[h!]
\centering
		\caption{\footnotesize{Comparison of Model Fit and Computational Efficiency in the Case Study}}
		\label{tab:LIKCASE}
		\resizebox{1\textwidth}{!}{
\begin{tabular}{ccccccccccccccc}
\hline 
\textbf{} & \multicolumn{4}{c}{\textbf{(-)Loglikelihood}} & \textbf{} & \multicolumn{4}{c}{\textbf{Estimation time (seconds)}} & \textbf{} & \multicolumn{4}{c}{\textbf{Loglikelihood function counts}} \\ \cline{1-5} \cline{7-10} \cline{12-15} 
\textbf{Draws} & \textbf{Halton} & \textbf{MLHS} & \multicolumn{2}{c}{\textbf{DQ}} & \textbf{} & \textbf{Halton} & \textbf{MLHS} & \multicolumn{2}{c}{\textbf{DQ}} & \textbf{} & \textbf{Halton} & \textbf{MLHS} & \multicolumn{2}{c}{\textbf{DQ}} \\ \cline{1-5} \cline{7-10} \cline{12-15} 
\textbf{} & \textbf{} & \textbf{} & \textbf{r=5} & \textbf{r=6} & \textbf{} & \textbf{} & \textbf{} & \textbf{r=5} & \textbf{r=6} & \textbf{} & \textbf{} & \textbf{} & \textbf{r=5} & \textbf{r=6} \\ \cline{1-5} \cline{7-10} \cline{12-15} 
100 & 5470.3 & 5470.6 & 5445.5 &  &  & 46 & 47 & 48 &  &  & 87 & 94 & 96 &  \\
200 & 5459.9 & 5462.6 & 5491.5 & 5445.2 &  & 90 & 86 & 87 & 85 &  & 84 & 86 & 84 & 84 \\
300 & 5455.7 & 5457.8 & 5457.3 & 5437.2 &  & 138 & 134 & 142 & 136 &  & 89 & 89 & 93 & 89 \\
1000 & 5453.0 & 5453.3 &  &  &  & 449 & 449 &  &  &  & 85 & 86 &  &\\
\hline
\end{tabular}}
\end{table}		

Table \ref{tab:PARCASE} shows the average parameter estimates and z-scores for selected QMC draws and DQ nodes. The mean estimates of all three methods are similar and in fact, z-score values are also in a similar range. The Cholesky components (e.g., L22 and L33) which are statistically significant in QMC remains significant in DQ and as expected, corresponding point estimates are also more stable across the considered draws.

\begin{table}[h!]
		\centering
		\caption{\footnotesize{Comparison of Estimates and Standard Errors in the Case Study}}
		\label{tab:PARCASE}
		\resizebox{1\textwidth}{!}{
\begin{tabular}{lccccccccccccccccc}
\hline

\textbf{} & \multicolumn{8}{c}{\textbf{Estimates}} & \textbf{} & \multicolumn{8}{c}{\textbf{z-scores}} \\ \cline{1-9} \cline{11-18} 
\textbf{} & \multicolumn{2}{c}{\textbf{Halton}} & \textbf{} & \multicolumn{2}{c}{\textbf{MLHS}} & \textbf{} & \multicolumn{2}{c}{\textbf{DQ (r=6)}} & \textbf{} & \multicolumn{2}{c}{\textbf{Halton}} & \textbf{} & \multicolumn{2}{c}{\textbf{MLHS}} & \textbf{} & \multicolumn{2}{c}{\textbf{DQ (r=6)}} \\ \cline{1-3} \cline{5-6} \cline{8-9} \cline{11-12} \cline{14-15} \cline{17-18} 
\textbf{Covariates} & \textbf{200} & \textbf{1000} & \textbf{} & \textbf{200} & \textbf{1000} & \textbf{} & \textbf{200} & \textbf{300} & \textbf{} & \textbf{200} & \textbf{1000} & \textbf{} & \textbf{200} & \textbf{1000} & \textbf{} & \textbf{200} & \textbf{300} \\ \cline{1-3} \cline{5-6} \cline{8-9} \cline{11-12} \cline{14-15} \cline{17-18} 
\textbf{Mean} &  &  &  &  &  &  &  &  &  &  &  &  &  &  &  &  &  \\ \cline{1-3} \cline{5-6} \cline{8-9} \cline{11-12} \cline{14-15} \cline{17-18} 
OVTT/100 (minute) & -1.5 & -1.6 &  & -1.7 & -1.6 &  & -1.8 & -1.5 &  & -3.1 & -3.0 &  & -3.4 & -3.0 &  & -4.0 & -3.0 \\
IVTT/100 (minute) & -10.8 & -11.0 &  & -10.8 & -11.0 &  & -11.0 & -11.3 &  & -14.9 & -14.5 &  & -14.8 & -14.5 &  & -15.3 & -14.4 \\
Trip Cost/10 (\$) & -3.1 & -3.1 &  & -3.1 & -3.1 &  & -3.5 & -3.4 &  & -15.3 & -14.4 &  & -15.1 & -14.3 &  & -16.0 & -15.4 \\
Electric? & -0.4 & -0.4 &  & -0.4 & -0.4 &  & -0.4 & -0.4 &  & -5.9 & -5.7 &  & -5.8 & -5.7 &  & -5.8 & -5.7 \\
Automation? & -0.5 & -0.5 &  & -0.5 & -0.5 &  & -0.5 & -0.5 &  & -6.5 & -6.4 &  & -6.5 & -6.4 &  & -6.3 & -6.2 \\ \cline{1-3} \cline{5-6} \cline{8-9} \cline{11-12} \cline{14-15} \cline{17-18} 
\textbf{Cholesky components} &  &  &  &  &  &  &  &  &  &  &  &  &  &  &  &  &  \\ \cline{1-3} \cline{5-6} \cline{8-9} \cline{11-12} \cline{14-15} \cline{17-18} 
L11 & 0.7 & 0.0 &  & 0.4 & 0.3 &  & 0.2 & -1.3 &  & 1.0 & -0.1 &  & 0.7 & 0.5 &  & 0.1 & -1.6 \\
L21 & 0.5 & 0.1 &  & 1.1 & 0.1 &  & 2.6 & 1.3 &  & 0.5 & 0.1 &  & 1.1 & 0.0 &  & 2.4 & 1.2 \\
L22 & 9.7 & 10.1 &  & 9.1 & 10.6 &  & 9.9 & 7.9 &  & 12.0 & 11.9 &  & 11.3 & 12.4 &  & 11.3 & 10.4 \\
L31 & 0.3 & 0.0 &  & 0.3 & 0.1 &  & 1.0 & 0.2 &  & 1.6 & 0.1 &  & 1.3 & 0.5 &  & 5.1 & 0.1 \\
L32 & 1.4 & 1.4 &  & 1.4 & 1.5 &  & 2.0 & 2.5 &  & 8.6 & 6.3 &  & 8.0 & 6.8 &  & 8.7 & 12.2 \\
L33 & 2.9 & 3.2 &  & 2.9 & 2.8 &  & 3.1 & 2.5 &  & 13.5 & 13.2 &  & 13.1 & 12.0 &  & 14.4 & 10.0 \\
L41 & -0.1 & 0.0 &  & 0.0 & 0.0 &  & 0.1 & 0.3 &  & -0.6 & 0.0 &  & -0.3 & -0.4 &  & 1.1 & 2.7 \\
L42 & 0.2 & 0.2 &  & 0.2 & 0.2 &  & 0.1 & 0.3 &  & 1.4 & 1.6 &  & 1.6 & 1.6 &  & 1.0 & 2.3 \\
L43 & 0.5 & 0.4 &  & 0.5 & 0.4 &  & 0.5 & 0.4 &  & 4.0 & 3.0 &  & 4.6 & 2.9 &  & 5.0 & 3.0 \\
L44 & -0.1 & 0.2 &  & 0.1 & -0.1 &  & 0.1 & 0.2 &  & -0.5 & 1.1 &  & 0.8 & -0.9 &  & 1.7 & 1.7 \\
L51 & -0.1 & 0.0 &  & 0.0 & 0.0 &  & 0.1 & 0.2 &  & -0.6 & 0.0 &  & -0.4 & -0.3 &  & 0.4 & 1.9 \\
L52 & 0.1 & 0.2 &  & 0.2 & 0.2 &  & 0.1 & 0.2 &  & 1.0 & 1.1 &  & 1.2 & 1.1 &  & 0.5 & 1.9 \\
L53 & 0.5 & 0.4 &  & 0.5 & 0.4 &  & 0.6 & 0.4 &  & 3.7 & 2.6 &  & 4.2 & 2.6 &  & 5.2 & 2.6 \\
L54 & -0.1 & 0.1 &  & 0.0 & -0.1 &  & -0.2 & 0.0 &  & -0.3 & 0.5 &  & 0.2 & -0.5 &  & -1.3 & 0.2 \\
L55 & 0.0 & 0.0 &  & 0.0 & 0.0 &  & 0.2 & -0.1 &  & 0.1 & 0.0 &  & 0.1 & 0.1 &  & 1.2 & 0.0 \\
\hline
\multicolumn{11}{l}{\textbf{Note:} OVTT: out-of-vehicle (walking and waiting) time, IVTT: in-vehicle travel time.} \\
\end{tabular} 
}
\end{table}

\section{Conclusions} \label{sec:con}
In this study, we have proposed the use of designed quadrature (DQ) to approximate multi-dimensional integrals in maximum simulated likelihood estimation of discrete choice models. We have compared performance of DQ with workhorse QMC methods in a Monte Carlo and an empirical case study. 

Whereas traditional sparse grid quadrature methods suffer from the problem of complex-valued loglikelihood due to negative weights, DQ could estimate MMNL smoothly for DGPs with varying covariance structures, thanks to positivity of weights. The simulation study confirmed that DQ requires too fewer function evaluations than QMC if the variance-covariance matrix is diagonal. Even in DGPs with non-diagonal matrices and full covariance structures, DQ always outperforms MLHS in terms of model fit and parameter recovery when the quadrature rule is generated on higher order polynomial subspaces. 

In sum, features like positivity of weights, computational efficiency due to fewer function evaluations, and easy implementation due to reusability of quadrature rules make DQ an attractive alternative to QMC methods. As a future work, we plan to test the performance of DQ in other discrete choice models (e.g., multinomial probit, and semi-parametric logit models). Furthermore, to ensure better performance of DQ over QMC, the key question is: for a given dimension and number of draws, on what maximum order of polynomial subspaces, DQ rule can be generated?  Thus, also as future work, taking advantage of the re-usability feature of DQ we plan to create software that can store the DQ rules on the highest possible order for commonly encountered dimensions, weight functions, and the number of nodes. With said software, DQ is as easy to use as any other QMC method, but with better performance. In other words, similar to QMC methods, the user would just need to choose the number of draws for the given dimension and software can provide the best DQ rule.  

\section*{Acknowledgements}

The authors are thankful to the National Science Foundation CAREER Award CBET-1253475 for financially supporting this research. They also thank Kenneth Train for sharing his initial thoughts on the applications of digital nets.   

\section*{Author contribution statement}

PB: conception and design, method derivation, method implementation,  data preparation and analysis, manuscript writing and editing; \\
VK: conception and design, method derivation, method implementation, manuscript editing; \\
AG: resources, supervision, manuscript editing; \\ 
RAD: resources, supervision, manuscript editing; \\
SL: resources, supervision, manuscript editing. \\

\newpage
\bibliographystyle{apalike}
\bibliography{reference.bib}

\begin{thebibliography}{}

\bibitem[Abay, 2015]{abay2015evaluating}
Abay, K.~A. (2015).
\newblock Evaluating simulation-based approaches and multivariate quadrature on
  sparse grids in estimating multivariate binary probit models.
\newblock {\em Economics Letters}, 126:51--56.

\bibitem[Askey, 1975]{askey1975orthogonal}
Askey, R. (1975).
\newblock {\em Orthogonal polynomials and special functions}, volume~21.
\newblock Siam.

\bibitem[Bhaduri and Graham-Brady, 2018]{bhaduri2018efficient}
Bhaduri, A. and Graham-Brady, L. (2018).
\newblock An efficient adaptive sparse grid collocation method through
  derivative estimation.
\newblock {\em Probabilistic Engineering Mechanics}, 51:11--22.

\bibitem[Bhat, 2001]{bhat2001quasi}
Bhat, C.~R. (2001).
\newblock Quasi-random maximum simulated likelihood estimation of the mixed
  multinomial logit model.
\newblock {\em Transportation Research Part B: Methodological}, 35(7):677--693.

\bibitem[Bhat, 2003]{bhat2003simulation}
Bhat, C.~R. (2003).
\newblock Simulation estimation of mixed discrete choice models using
  randomized and scrambled halton sequences.
\newblock {\em Transportation Research Part B: Methodological}, 37(9):837--855.

\bibitem[Bhat, 2011]{bhat2011maximum}
Bhat, C.~R. (2011).
\newblock The maximum approximate composite marginal likelihood (macml)
  estimation of multinomial probit-based unordered response choice models.
\newblock {\em Transportation Research Part B: Methodological}, 45(7):923--939.

\bibitem[Bhat, 2015]{bhat2015new}
Bhat, C.~R. (2015).
\newblock A new generalized heterogeneous data model (ghdm) to jointly model
  mixed types of dependent variables.
\newblock {\em Transportation Research Part B: Methodological}, 79:50--77.

\bibitem[Brumm and Scheidegger, 2017]{brumm2017using}
Brumm, J. and Scheidegger, S. (2017).
\newblock Using adaptive sparse grids to solve high-dimensional dynamic models.
\newblock {\em Econometrica}, 85(5):1575--1612.

\bibitem[Cagnone and Bartolucci, 2017]{cagnone2017adaptive}
Cagnone, S. and Bartolucci, F. (2017).
\newblock Adaptive quadrature for maximum likelihood estimation of a class of
  dynamic latent variable models.
\newblock {\em Computational Economics}, 49(4):599--622.

\bibitem[Davis and Rabinowitz, 2007]{davis2007methods}
Davis, P.~J. and Rabinowitz, P. (2007).
\newblock {\em Methods of numerical integration}.
\newblock Courier Corporation.

\bibitem[Dick and Pillichshammer, 2014]{dick2014discrepancy}
Dick, J. and Pillichshammer, F. (2014).
\newblock Discrepancy theory and quasi-monte carlo integration.
\newblock In {\em A Panorama of Discrepancy Theory}, pages 539--619. Springer.

\bibitem[Geweke et~al., 1994]{geweke1994alternative}
Geweke, J., Keane, M., and Runkle, D. (1994).
\newblock Alternative computational approaches to inference in the multinomial
  probit model.
\newblock {\em The review of economics and statistics}, pages 609--632.

\bibitem[Golub and Welsch, 1969]{golub1969calculation}
Golub, G.~H. and Welsch, J.~H. (1969).
\newblock Calculation of gauss quadrature rules.
\newblock {\em Mathematics of computation}, 23(106):221--230.

\bibitem[Goos and Mylona, 2018]{goos2018quadrature}
Goos, P. and Mylona, K. (2018).
\newblock Quadrature methods for bayesian optimal design of experiments with
  nonnormal prior distributions.
\newblock {\em Journal of Computational and Graphical Statistics},
  27(1):179--194.

\bibitem[Heiss, 2010]{heiss2010panel}
Heiss, F. (2010).
\newblock The panel probit model: Adaptive integration on sparse grids.
\newblock In {\em Maximum simulated likelihood methods and applications}, pages
  41--64. Emerald Group Publishing Limited.

\bibitem[Heiss and Winschel, 2008]{heiss2008likelihood}
Heiss, F. and Winschel, V. (2008).
\newblock Likelihood approximation by numerical integration on sparse grids.
\newblock {\em journal of Econometrics}, 144(1):62--80.

\bibitem[Hess et~al., 2006]{hess2006use}
Hess, S., Train, K.~E., and Polak, J.~W. (2006).
\newblock On the use of a modified latin hypercube sampling (mlhs) method in
  the estimation of a mixed logit model for vehicle choice.
\newblock {\em Transportation Research Part B: Methodological}, 40(2):147--163.

\bibitem[Jakeman and Narayan, 2018]{jakeman2018generation}
Jakeman, J.~D. and Narayan, A. (2018).
\newblock Generation and application of multivariate polynomial quadrature
  rules.
\newblock {\em Computer Methods in Applied Mechanics and Engineering},
  338:134--161.

\bibitem[Keshavarzzadeh et~al., 2018]{keshavarzzadeh2018numerical}
Keshavarzzadeh, V., Kirby, R.~M., and Narayan, A. (2018).
\newblock Numerical integration in multiple dimensions with designed
  quadrature.
\newblock {\em SIAM Journal on Scientific Computing}, 40(4):A2033--A2061.

\bibitem[Liu et~al., 2018]{liu2018framework}
Liu, Y., Bansal, P., Daziano, R., and Samaranayake, S. (2018).
\newblock A framework to integrate mode choice in the design of
  mobility-on-demand systems.
\newblock {\em arXiv preprint arXiv:1805.06094}.

\bibitem[Ma and Zabaras, 2009]{ma2009adaptive}
Ma, X. and Zabaras, N. (2009).
\newblock An adaptive hierarchical sparse grid collocation algorithm for the
  solution of stochastic differential equations.
\newblock {\em Journal of Computational Physics}, 228(8):3084--3113.

\bibitem[Munger et~al., 2012]{munger2012estimation}
Munger, D., L'Ecuyer, P., Bastin, F., Cirillo, C., and Tuffin, B. (2012).
\newblock Estimation of the mixed logit likelihood function by randomized
  quasi-monte carlo.
\newblock {\em Transportation Research Part B: Methodological}, 46(2):305--320.

\bibitem[Patil et~al., 2017]{patil2017simulation}
Patil, P.~N., Dubey, S.~K., Pinjari, A.~R., Cherchi, E., Daziano, R., and Bhat,
  C.~R. (2017).
\newblock Simulation evaluation of emerging estimation techniques for
  multinomial probit models.
\newblock {\em Journal of choice modelling}, 23:9--20.

\bibitem[Richard and Zhang, 2007]{richard2007efficient}
Richard, J.-F. and Zhang, W. (2007).
\newblock Efficient high-dimensional importance sampling.
\newblock {\em Journal of Econometrics}, 141(2):1385--1411.

\bibitem[Ryu and Boyd, 2015]{ryu2015extensions}
Ryu, E.~K. and Boyd, S.~P. (2015).
\newblock Extensions of gauss quadrature via linear programming.
\newblock {\em Foundations of Computational Mathematics}, 15(4):953--971.

\bibitem[S{\'a}ndor and Train, 2004]{sandor2004quasi}
S{\'a}ndor, Z. and Train, K. (2004).
\newblock Quasi-random simulation of discrete choice models.
\newblock {\em Transportation Research Part B: Methodological}, 38(4):313--327.

\bibitem[Smolyak, 1963]{smolyak1963quadrature}
Smolyak, S.~A. (1963).
\newblock Quadrature and interpolation formulas for tensor products of certain
  classes of functions.
\newblock In {\em Doklady Akademii Nauk}, volume 148, pages 1042--1045. Russian
  Academy of Sciences.

\bibitem[Train, 2009]{train2009discrete}
Train, K.~E. (2009).
\newblock {\em Discrete choice methods with simulation}.
\newblock Cambridge university press.

\bibitem[Yu et~al., 2010]{yu2010comparing}
Yu, J., Goos, P., and Vandebroek, M. (2010).
\newblock Comparing different sampling schemes for approximating the integrals
  involved in the efficient design of stated choice experiments.
\newblock {\em Transportation Research Part B: Methodological},
  44(10):1268--1289.

\end{thebibliography}

\end{document}